\magnification=1200
\nopagenumbers
\parskip=10pt
\vskip .7truein
\centerline{\bf{ Equivalence of Constrained Models}}
\vskip 1.0truein
\centerline {M.Horta\c csu*, K. \" Ulker*}
\vskip .3truein
\centerline{ * Physics Dept., Faculty of Science and Letters}
\vskip .1truein
\centerline{ I.T.U., 80626 Maslak, Istanbul, Turkey}
\vskip .3truein
\vskip 1.7truein
Abstract:  We study two constrained scalar models.  While there seems to be
equivalence when the partially integrated Feynman path integral is expanded 
graphically, the dynamical behaviours of the two models are different
when quantization is done using Dirac constraint analysis.
\vfill\eject
\baselineskip=18pt
\footline={\centerline{\folio}}
\pageno=1
\noindent INTRODUCTION

\noindent
There are many models in the literature which claim that they are equivalent 
to Quantum Chromodynamics (QCD) in some sense.  We can site a recent work by
Hasenfratz and Hasenfratz $^{/1}$, or the work of Akdeniz et al $^{/2}$
among many papers concerning this topic.  We mention these two papers
since in these papers the starting lagrangians look very different,
although the effective lagrangians obtained after some manipulations are made,
are exactly the same.  There are many other papers in the same spirit.
One example is reference 3, which , in some sense, gave rise to reference 1.
Another example is reference 4, which was actually a pioneering paper in
this endeavor of finding models whose effective lagrangian looks like the
standard model.

It was always a puzzle to us how similar manipulations made on very 
differently looking Lagrange functions resulted in completely the same 
effective lagrangian.  In this note we try to investigate this phenomena  
using scalar models.  We think that our results in the scalar case may give 
additional information on this phenomenum.

We will use two constrained scalar models.  In reference 1, the authors
imposed the constraint $ J^{f}_{\mu}=\overline{ \Psi}i \gamma_{\mu} 
\tau _a \psi =0 $ on the 
fields of the free spinor lagrangian.  To resemble this model here we first 
study the case where we write a lagrangian which is essentially
equivalent to $ L =  {1 \over {2} } {\partial _{\mu}} \phi
{\partial ^ {\mu}} \phi $ .  We will impose a different constraint, though, 
since a constraint $\phi \partial_{\mu} \phi=0 $ results in a truly trivial 
model. We can also introduce inner symmetry to the theory and make a $O(N)$ 
model along similar lines.
For the time being we do not pursue this .

The authors in reference 2 impose the condition that their current 
$ J^{f}_{\mu} $ equals product of vector fields instead of zero,
$ J^{f}_{\mu} = A_{\mu} A^{2} $, which differs from the constraint
used in reference 1.
This complicates the problem, but all of additional fields introduced to
the model decouple and at the end only one vector field survives.  The
propagator for this field, and only for this field, is generated in the
one loop calculation.  At this point the resulting effective theory looks
exactly like that of reference 1.

We have doubts whether these two models are actually equivalent to QCD 
in all aspects.  One may refer to an old work of Wilson $^{/5}$ and to a 
more recent work of Zinn-Justin $^{/6}$, and using the calculations made
in reference 7, claim that these two models are actually examples of 
trivial models $^{/8} $. 
  
We will not dwell on these points here.  We will only investigate in what 
sense two models are  equivalent when the effective lagrangians derived 
from them seem so.  In the next section we present two constrained scalar
models.
We get a theory which is totally trivial if we impose the current
made out of scalar fields equal to zero, the analogous case as given in
reference 1.  We instead use two models where
the current is equal to one and two auxiliary fields, thus introducing
eight and sixteen new degrees of freedom respectively plus constraints that
will eliminate these .
We study the Dirac bracket relations satisfied by the respective fields  .
We see that the new introduced vector fields via the constraint equations
somehow replace the canonical momentum of the scalar field.

In Section III we derive effective lagrangians for these two cases and show 
why do they seem to be equivalent  on this level.  We end with some remarks.

\bigskip
\noindent   
{\bf {II. Quantization of the Models using Dirac Constraint Analysis}}

\noindent
II. A

\noindent
We start with 
$$ L_A = {1\over 2}{\partial_{\mu}} \phi {\partial ^{\mu}} \phi 
   +ig\lambda_{\mu} \phi \partial^{\mu} \phi 
   + {1\over {2}} g^2 \phi^2 \lambda^2.
\eqno{1} $$
We are in four dimensional Minkowski space and $\mu$ takes the values zero to 
three.
Here $\lambda_{\mu}$ is an auxiliary field with no kinetic term. 
$g$ is a coupling constant.  

The equations of motion are
$$ \partial_{\mu} \partial ^{\mu} \phi +ig \partial _{\mu} ( \lambda ^{\mu} 
\phi)=
ig \lambda_{\mu} \partial^{\mu} \phi +g^2 \lambda^2 \phi ,\eqno {2} $$
$$ ig \phi \partial _{\mu} \phi = -g^2 \lambda_{\mu} \phi^2 \eqno {3} $$
which can be shown to be equivalent to
$$\partial_{\mu} \partial ^{\mu} \phi =0 .  \eqno {4} $$

In this calculation we use the methods given in Dirac's book $^{/9}$.
The canonical momenta are
$$ \pi_{\phi} = ({\partial_0} + ig\lambda_0) \phi \eqno{5a} $$
and
$$\pi_{\lambda_{\mu}} =0 \eqno {5b} $$
which gives us four primary constraints.

\noindent
The canonical hamiltonian is
$$H= {1\over 2} ( \pi_{\phi}-ig\lambda_0 \phi )^2 + 
ig \lambda_i \phi {\partial _i} \phi +{1\over 2}
{\partial_i} \phi {\partial_i} \phi -{1\over2}g^2 \lambda^2 \phi^2 .\eqno {6} $$
We get secondary constraints when we set the Poisson bracket of $ H $
with $ \pi_{\lambda _{\mu}}$
equal to zero.
$$ K_\mu= {d \over{dt}} 
\pi_{\lambda_{\mu}}= 
(ig \phi \pi_\phi +g^2\lambda_0 \phi^2) g_{0 \mu} 
-ig\phi {\partial _i} \phi g_{i \mu} +g^2\lambda_{\mu} \phi^2 =0. \eqno{7} $$
We take 
$$ H_{E} = H + c_{\mu} \pi _{\lambda _{\mu}} \eqno {8} $$
and further calculate
$$ {d \over {dt}} K_{\mu} = [ H_{E} , K_{\mu} ] \eqno {9} $$
where square brackets mean Poisson brackets.
Note that  $\lambda_0$ appears in $K_{0}$, and the $\lambda_{i}$ in $K_i$.  
We get one equation with
$c_0$ when $ [ H_E, K_0] $ is calculated, which fixes the value of $c_0$ and
does not give additional constraints on the system.  $[H_E, K_i] $ give us 
equations which fix
$c_i$ .   We do not get any additional constraints.

We can calculate the Poisson brackets between the different constraints.
$$[\pi_{\lambda_0}, K_0]=  2g^2\phi^2 g_{00},\quad 
[\phi_{\lambda_i},K_j]=-g^2\phi^2 g_{ij},$$
$$ [K_0,K_i]=g^2 \phi \partial_i \phi-2ig^3\lambda_i\phi^2 $$
all the other brackets of the constraints with each other are zero.  We see 
that all these brackets are second class.
We calculate Dirac brackets between different fields.
$$[\phi(x),\lambda_0(y)]^{D} = {i\over {2g\phi}} \delta^3 (x-y),\eqno {10a} $$
$$[\phi(x), \lambda_i]=0 ,\eqno {10b}$$
which shows that $\lambda_0$ is like $\pi_{\phi}$, and $\lambda_i$ decouples.
We can set $\lambda_i$ equal to zero .
\bigskip
\noindent
II.B

\noindent
We propose another model where
$$ L_B = {1\over 2} {\partial _{\mu}} \phi {\partial ^{\mu}} \phi 
+ig (\lambda_{\mu}+A_{\mu})\phi {\partial ^{\mu}}\phi
-g A_{\mu}  \lambda^{\mu} A^2 \eqno {11} $$
The primary constraints are
$$ \pi_{A_{\mu}}=0 \eqno {12} $$
$$ \pi_{\lambda_{\mu}} =0 \eqno {13} $$
The hamiltonian reads
$$ H = {1\over 2} ( \pi_{\phi} -ig(\lambda_0+A_0)\phi)^2
+ {1\over 2} {\partial _i} \phi {\partial _i} \phi 
+ ig(\lambda_i+A_i) \phi {\partial _i} \phi +g \lambda_{\mu} A^{\mu} A^2 
\eqno {14} $$
where
$$ \pi_{\phi}={\partial_0} \phi +ig(\lambda_0+A_0)\phi \eqno{15} $$
$$ H_E= H + c_1^{\mu} \pi _{A_\mu}+ c_2 ^{\mu} \pi _{\lambda _\mu}\eqno{16} $$
Secondary constraints
$$[H_E,\pi_{A_{\mu}}, ] = Q^{1}_{\mu} =0 \eqno {17} $$
$$[H_E,\pi_{\lambda_{\mu}}]= Q^2_{\mu} =0 \eqno{18} $$
are given as 
$$Q^{1}_{\mu}=ig \phi(\pi_{\phi}-ig(\lambda_0+A_0)\phi)g_{\mu 0}
-ig\phi {\partial _i} \phi g_{i \mu} 
-g\lambda_{\mu} A^2 -2g\lambda_{\nu}A^{\nu} A_{\mu} \eqno {19} $$
$$Q^{2}_{\mu} =ig\phi(\pi_{\phi}-ig(\lambda_0+A_0)\phi ) g_{\mu 0} -ig\phi 
{\partial _i \phi g_{i \mu}-g A_{\mu}} A^2 \eqno {20} $$
We see that the system is closed, since the Poisson brackets of $Q^{1}_{\mu}, Q^{2}_{\mu}$
with $H_E$ involve eight coupled equations for $c^{1}_{\mu}$ and 
$c^{2}_{\mu}$.
We get no further constraints.

When we calculate the Poisson brackets of the constraints with each other,
we see that they are all of second class.  We have sixteen second class 
constraints and eighteen degrees of freedom.  We have traded some of our 
fields in terms of others, but we did not change the 
number of independent variables.

Now we can calculate the Dirac brackets between different fields.  We are 
particularly
interested  in the brackets between $\phi$ and $A_{\mu}, \lambda_{\mu}$,
since the Poisson brackets between the same fields are zero.
The effect of the constraints in the system are reflected to the Dirac 
brackets; hence, they do not vanish when thay are taken between $\phi$ and 
the auxiliary fields.
We give  below the result of some sample calculations.
$$ [\phi (x), A_1 (y)]^{D} \phi (y) ={{i 
\phi^2(-2A_0 A_1) [g\phi^2(A^2-2{\bf{A^2}})-3A^4] } 
\over{\Delta}}\delta^{(3)} (x-y), \eqno {21a} $$
$$ [\phi (x), A_0 (y)]^{D} \phi (y) ={{i   \phi^2(A^2-
2{\bf{A^2}}) [g\phi^2 (A^2- 2{\bf{A^2}})-3A^4] } \over {\Delta}} 
\delta^{(3)} (x-y),\eqno {21b} $$
$$ [\phi (x), \lambda_0 (y)]^{D} \phi (y) = {{i \phi^{2} 
(A^2-2{\bf{A^{2}}})[2g\phi^2(A^2-2{\bf{A^2}})-3A^4-4A^2 \lambda_{\mu} 
A^{\mu} ] } \over{\Delta}} \delta^{(3)} (x-y), \eqno {21c} $$
$$ [\phi (x), \lambda_2 (y)]^D \phi (y)= 2i { { \phi^2[-3A_0A_2(A^4-
4A^2A_{\mu} \lambda ^{\mu})-3A^4(A_0 \lambda_2+A_2\lambda_0)]} 
\over {\Delta}} \delta^{(3)} (x-y) \eqno {21d}        $$
where 
$$\Delta = \left[ g^2\phi^4(A^2-2 {\bf{A ^2}})^2-(6A^2+4\lambda_{\mu} 
A^{\mu})(A^2-2{\bf{A^2}})A^2 g \phi^2+9A^8 \right]   \eqno{22} $$
Here $\bf{A^2}$ means the three vector $A$ squared.

Upon quantization we see that $A_{\mu}$ and $\lambda_{\mu}$ seem to contain 
part of $\pi_{\phi}$.  We expect only $\pi_{\phi}$ to have nonzero 
commutation relations with $\phi$ and in this model both $A_{\mu}$ and 
$\lambda_{\mu}$ also will have non zero commutations 
with $\phi$.  The constraints $Q^{1}_{\mu} =0$ and $Q^{2}_{\mu} =0$ 
relates $\pi_{\phi}$ to these fields.

The model we studied seems to be considerably different from the one studied 
in the first  Section.  The fields in the model has
non zero Dirac brackets; so, we can not set them equal to zero, as in the
previous model.
The space components of the vector fields do not decouple
and can not be set to zero.

Note that in both of these models the degrees of freedom is two.  In Model A 
we start with ten degrees of freedom, two for the $\phi$ and eight for the 
$\lambda$ field and their respective momenta.  Eight constraints reduce 
these to two.  In Model B we start with eighteen degrees of freedom since we 
have two vector particles.
Sixteen constraint equations reduce this number to two.
As far as the equations of motion are considered these two models do not seem 
to be alike.

\noindent
{\bf{III. { Feynman Rules using the Path Integral}}}

\noindent
Here we study the two models using Feynman diagram expansions of the path 
integral after the integral is partially integrated.  We start by studying 
model A, then contrast our results with that of Model B.

\noindent
III.A
\noindent
Here the path integral is written as
$$Z= \int d\phi d\pi_{\phi} d\lambda_{\mu} d\pi_{\lambda_{\mu}}
\delta (\pi_{\lambda_\mu} ) \delta(K_{\mu}) det M_{\mu \nu} \exp{i S}, 
\eqno {23}$$
where
$$ M_{\mu \nu} = {{\partial K_{\mu}}\over {\partial \lambda_{\nu}}} ,
\eqno {24a}$$
$$S=\int d^4 x [\pi_{\phi} \partial_0 \phi + \pi_{\lambda_{\mu}} 
\partial_0 \lambda _{\mu} -H_E]. \eqno{24b}$$
We write the Dirac delta functions in the integral form, introducing new 
variables $A_{\mu}$
and express the determinant in the exponential form using ghost fields.
$$ \delta (K_{\mu}) = {1\over {2\pi}} \int dA_{\mu} e^{-i A_{\mu} K^{\mu}}, $$
$$ det M_{\mu \nu} = \int dc_\mu^+ dc_\nu e^{i c_{\mu}^{+} M^{\mu \nu} 
c_{\nu}} .$$
The integrations  over the momenta and $\phi$ are performed easily  and we 
end up with
$$ Z= N \int dA_{\mu} d\lambda_{\nu}dc_{\alpha}^{+} dc_{\beta}
e^{-{i\over2} tr log [-\partial_{\mu} \partial^{\mu}  
+igN_{\mu} \partial^{\mu}-ig \partial_{\mu} N^{\mu}
+g^2({1\over 2} \lambda^2 -A_{\mu} \lambda ^{\mu} -{1\over 2} A_0^2
+c_0^{+} c_0 +c_{\mu}^{+}c^{\mu})] }.\eqno {25} $$ where we
define $N_{\mu} = \lambda_{\mu} -A_{\mu} $.
We can calculate the inverse propagator, $D^{-1}_{\mu \nu}$ for the $N _{\mu} $
field by taking two derivatives of eq. (25) with respect to the $N_{\mu}$
field.  In the momentum representation we get
$$ D^{-1}_{\mu \nu} (q) =- g^2 \int {d^4 p \over {(2\pi)^4}} ({p_{\mu}+
q_{\mu}) (p_{\nu}+2q_{\nu}) \over { p^2 (p+q)^2}}  $$
$$ = -g^2 {\Gamma (\epsilon)\over{6 (4\pi)^2}} \left( g_{\mu \nu}q^2 - 
10 q_{\mu} q_{\nu} \right) \eqno {26} $$
which looks like the massless vector boson propagator, at least in a 
particular gauge.
Note that all the components of the vector field have non-zero propagation .

All other fields have
zero propagators if we use dimensional regularization.
Here we set $\int d^4p {1\over {p^2}} =0 $.  When we drop all the fields with
zero propagators we end up with
$$S_{eff} = -{1\over 2} Tr log ( -{\partial}^2+igN_{\mu} {\partial}^{\mu}-
ig\partial^{\mu} N_{\mu}). \eqno {27}$$
Upon expanding the logarithm we can evaluate 
the multi-point functions  for the $N^{\mu}$ fields.  Eq. 26 dictates a
necessary condition on the coupling constant $g$ , though, to have a well 
defined
expression for the propagator function given by this equation, which reads
$$g^2{\Gamma (\epsilon) \over { 6(4\pi)^2}} =1 .\eqno {28} $$
This condition makes the model asymptotically free in the ultraviolet regime.

\noindent
By taking all the non vanishing terms we see that  for the composite field 
$\lambda_{\mu}$ the effective lagrangian can be written as
$$L_{eff} = {1 \over 2} {\partial _\mu} N_{\nu} {\partial ^\mu}N _{\nu}
+{\partial _{\mu}} N_{\nu}{\partial ^{\nu}} N ^{\mu}
+g f^{\mu \nu \rho} N ^{\mu} N ^{\nu} N ^{\rho}
+g^2 V_{\mu \nu \rho \sigma} N ^{\mu} N ^{\nu} N ^{\rho} N ^{\sigma}. 
\eqno {29} $$
Here $f_{\mu \nu \rho} $ is proportional to momentum and Kronecker deltas and
$V_{\mu \nu \rho \sigma} $ is made out of Kronecker deltas.  Higher order 
functions, starting with the fifth
point function, drop with higher powers of $g$.  For example the five point
function goes as $g^5$.  They do not fit into this scheme of effective 
lagrangian and are calculated as loop corrections.

Here we calculated the Feynman rules for this model and showed that apart
from the restriction dictated by eq. 28, we get rules similar
to those as a gauge theory.
One can calculate physical processes using these rules and will get free 
parton model
results, as is the case in a similar model $^{/7}$   due to the restriction
dictated by eq. 28.   All the physical processes that involve interactions 
will involve powers of the coupling constant which goes to zero .  Any 
possible divergences due to loops will be canceled by the zeroes coming from 
extra powers of the coupling constant.  Only terms which do not involve any 
interactions are finite.  These terms are the same as those given in the free 
field case.

\bigskip
\noindent
III.B
\noindent
The path integral for Model B, in the hamiltonian formalism, is written as
$$\int d A_{\mu} d \pi_{A_{\mu}} d \pi_{\lambda{\mu}} d\lambda_{\mu} d \phi 
d \pi_{\phi}
\delta (\pi_{\lambda_{\mu}}) \delta (\pi_{A_{\mu}}) \delta (Q^{1}_{\nu})
\delta (Q^2_{\nu})(det M ) \exp {iS} \eqno {30} $$
Here 
$$ S= \int d^4 x [ \pi_\phi \partial _0 \phi 
+ \pi_{A_\mu}{\partial _0} A_\mu
+\pi_{\lambda_\mu}{\partial _0 }\lambda _\mu 
-{1\over 2} \left[\pi_{\phi}+ig(\lambda_0+A_0)\phi\right]^2
-{1\over 2} {\partial _i} \phi {\partial_{i}} \phi$$ 
$$-ig(\lambda_{i} + A_{i})\phi
\partial_{i} \phi -g\lambda_{\mu} A^{\mu} A^2 ] \eqno {31} $$
$$Q^{1}_{\mu} = \phi (\pi_{\phi}-(\lambda_{0} +A_{0})\phi ) g_{\mu 0}
-\phi {\partial _{i}} \phi g_{i \mu} -\lambda_{\mu} A^2 -2\lambda_ {\nu}
A^{\nu} A_{\mu} \eqno {32} $$
$$Q^{2}_{\mu} = \phi (\pi_{\phi}-(\lambda_{0} + A_{0})\phi) g_{\mu  0}
-\phi {\partial _{i} }\phi g_{i \mu} -A_{\mu} A^2 \eqno {33} $$
M is a eight by eight matrix whose entities are made out of derivatives 
of $ Q^{1}_{\mu}$ and $Q^{2}_{\mu}$ with respect to the fields $A_{\mu} $
and $\lambda_{\mu}$.

We can use the integral representation of the Dirac delta functions.
$$\delta (Q^{1}_{\mu})={1\over{2\pi}} \int dB_{\mu} \exp {-iB^{\mu} 
Q^{1}_{\mu}} \eqno {34} $$
$$\delta (Q^{2}_{\mu})={1\over{2\pi}} \int dE_{\mu} \exp {-iE^{\mu} 
Q^{2}_{\mu}} \eqno {35} $$
Using ghost , i.e. Grassmann valued fields $c_{\mu},e_{\mu},c^{\dagger}_{\mu},
e^{\dagger}_{\mu}$, we can raise $det M $ to the exponential.
$$det M = \int dc^{\dagger}_{\mu} dc_{\nu} de^{\dagger}_{\sigma} de_{\rho} 
\exp{ iN} \eqno {36} $$
where
$$ N=(c^{\dagger}_{\mu} + e^{\dagger}_{\mu})(g^2\phi^2 g_{\mu 0} g_{\nu 0} )
( c_{\nu}+e_{\nu})+
c^{\dagger}_{\mu}(-2gA_{\mu} \lambda_{\nu}
-2gg_{\mu \nu} \lambda_{\kappa} A^{\kappa} -2g \lambda_{\mu} A_{\nu}
-2g\lambda_{\nu} A_{\mu}) c_{\nu}$$
$$+c^{\dagger}_{\mu} (-g^{\mu \nu} A^2 -2gA^{\mu} A^{\nu} ) e_{\nu}
+e^{\dagger}_{\mu} ( -gg^{\mu \nu}A^2-2g A^{\mu} A^{\nu} ) c_{\nu} 
\eqno {37} $$
When the momentum integrals are performed we get 
$$L_{eff} =i[{1\over 2} {\partial_{\mu}} \phi {\partial ^{\mu}} \phi
+ig G_{\mu}\phi {\partial ^{\mu}} \phi
-{{g^2}\over {2}} (B_0+E_0)^2 \phi^2 -g\lambda_{\mu} A^{\mu} A^2 +
gB^{\mu} \lambda _{\mu} A^2 $$
$$+2gB_{\mu} A^{\mu} \lambda_{\nu} A^{\nu} +gE{\mu} A_{\mu} A^2 
+ g^2{f^{\dagger}_0 }  \phi^2 f_{ 0}
+c^{\dagger}_{\mu} ( -gg^{\mu \nu}A^2 -2 gA^{\mu} A^{\nu} ) f_{\nu}$$
$$+f^{\dagger}_{\mu}(-gg^{\mu \nu} A^2 -2g A^{\mu} A^{\nu} ) e_{\nu}
+c^{\dagger}_{\mu}[ 2gA^2g^{\mu \nu}  -2 g\lambda^{\mu} A^{\nu}
-2gg^{\mu \nu} \lambda_{\kappa} A ^{\kappa}- 2 g A^{\mu}\lambda^{\nu}
 +4 gA^{\mu} A^{\nu} ] c_{\nu} \eqno{38} $$
Here $f_{\mu}=c_{\mu}+e_{\mu}$.
We set $G_{\mu}=A_{\mu}+\lambda_{\mu}-B_{\mu}-E_{\mu}$.

We perform the integration over $\phi$ and obtain
$$ S_{eff} = -{1\over {2}} {Tr log } [ -{\partial^2}
+igG_{\mu} {\partial^{\mu}}-ig\partial_{\mu}G
-{{g^2}\over{2}}(B_0+E_0)^2+g^2f_0^{\dagger} f_0]$$
$$+\int d^4 x \left[ -g\lambda^{\mu} A_{\mu}A^2+gB_{\mu}\lambda^{\mu}A^2
+2g\lambda^{\mu}A_{\mu} A_{\nu} B^{\nu}
+gE^{\mu}A_{\mu}A^2+
e^{\dagger}_{\mu}(-gA^2g^{\mu \nu} -2gA^{\mu}A^{\nu} )f_{\nu} \right. $$
$$\left. +f^{\dagger}_{\mu}(-gA^2g^{\mu \nu} -2gA^{\mu}A^{\nu})e_{\nu} 
 +c^{\dagger}_{\mu}(2gA^2 g^{\mu \nu} +4gA^{\mu} A^{\nu}
-2g\lambda^{\mu} A^{\nu}
-2g\lambda^{\nu} A^{\mu}-2g\lambda_{\rho } A^{\rho} g^{\mu \nu} ) c_{\nu} 
\right] \eqno {39} $$

Note that only $G_{\mu}$ propagates among all the fields given above.  To 
find the propagator we take two derivatives with respect to the respective 
fields.
$$ {\partial^2 S_{eff} \over {{\partial G_{\mu} (x)} {\partial G_{\nu} (y) 
}}}|_{0} = {-1\over {(2\pi)^4}} 
g^2 \int d^4 p {{(p^{\mu}+q^{\mu}) (p^{\nu}+2q^{\nu})} \over{p^2(p+q)^2}} 
\eqno {40} $$
Subscript zero on the derivative means that all the fields are put to zero 
after the differentiation is performed.

Note that this is the same expression for the propagator of the 
$\lambda_{\mu} $ field as given in eq.(26)
We also see that
$${\partial^2 S_{eff} \over {\partial B^2_{0} }}={\partial^2 S_{eff} 
\over {\partial E^2_{0}}}=
{\partial^2 S_{eff} \over {{\partial g^{\dagger}_{0}}{ \partial g_{0} }}}
={1\over {(2\pi)^4}} \int {d^4 p\over{p^2}} \eqno {41} $$ 
This expression is zero by dimensional regularization.All the other fields 
also have zero propagators since the effective lagrangian does not have any 
terms which are only bilinear in these fields.  All these terms involve 
quartic interactions of these fields.  When we
drop all the field with zero propagators we end up with
$$S''_{eff}= -{Tr log \over 2} (-{\partial ^2} -ig G_{\mu} \partial^{\mu}+
ig\partial _{\mu}G^{\mu} ) \eqno {42} $$
This is the same expression we found for Model A.  Therefore all the results 
obtained for Model A from this expression are also true for Model B.  
We can not differentiate Model A from Model B as far as perturbative 
expansion in terms of Feynman diagrams are concerned.
  
\bigskip
\noindent   {\bf {Conclusion}}

\noindent
Here we have studied two very dissimilar models which have the same Feynman 
expansions.  A complete constrained Hamiltonian analysis shows that the two 
models are  different.  One reason we have studied this problem is to be able 
to clarify the behaviour of many Nambu-Jona-Lasinio$^{/10,3}$like models 
which are claimed to be
similar to QCD $^{/11}$.  There are people $^{/12}$ who disagree with 
this equivalence.
The claim in reference 12 is that after an investigation of a lattice 
Nambu-Jona-Lasinio
model both by the Monte Carlo method and Schwinger-Dyson equations, studying
renormalization group flows in the neighborhood of the critical coupling
where the chiral symmetry breaking phase transition takes place, in no region of the
bare parameter space renormalizability of the model is found.
We propose that, in addition to the standard methods of looking at the
renormalization flow
and fixed point structure of two
models to show equivalence, their constrained analysis may be another check.
Still another method is to study the predictions of these models for 
different physical processes .  An old calculation $^{/7}$ and old paper 
$^{/5}$ seem to suggest that the Nambu-Jona-Lasinio type models may indeed 
be trivial at four dimensional space-time, perhaps like the $\phi^4$ model is.

\noindent
Acknowledgewment:  We are grateful to Prof. \"Omer Faruk Day\i\ for many 
very enlightening discussions. This work is partially supported by 
T\" UBITAK, the Scientific and Technical Research Council of Turkey .  
The work of M.H. is also supported by the Turkish Academy of Sciences .
\vfill\eject
\noindent
REFERENCES

\item {1.}  A.Hasenfratz and P.Hasenfratz, Phys. Lett. {\bf{297B}} (1992) 166.

\item {2.}  K.G.Akdeniz, M.Ar\i k,M.Horta\c csu, N.K.Pak, Phys.Lett. 
{\bf{124B}} (1983)79.

\item {3.}  A. Hasenfratz,P.Hasenfratz,K.Jansen,J.Kuti,Y.Shen, Nucl. Phys. 
B{\bf{365}} (1991)      .

\item {4.}  D.Amati, R.Barbieri, A.C.Davis and G.Veneziano, Phys. Lett.
{\bf{ 102B}} (1981) 408.
  
\item {5.}  K.G. Wilson, Phys. Rev. D{\bf{7}}, (1973) 2911.

\item {6.}  J.Zinn-Justin, Nucl. Phys.B {\bf{ 36}}7 (1991) 105.

\item {7.}  M.Ar\i k and M.Horta\c csu, J.Phys. G: Nucl. Phys. {\bf{9}} 
(1983) L119.

\item {8.}  M.Horta\c csu, Bull. Tech. Univ. Istanbul,{\bf{ 47}} (1994) 321;
            V.E.Rochev, J.Phys. A: Math. Gen.,{\bf{30}} (1997) 3671,
            V.E.Rochev and P.A. Saponov, " The four-fermion interaction in 
            D=2,3,4: a nonperturbative treatment", IHEP preprint, Moscow, 
            hep-th/9710006.

\item {9.}  P.A.M.Dirac, {\it {Lectures on Quantum Mechanics }}, Belfer 
Graduate School of Science, Yeshiva University, New York (1964)

\item {10.} Y.Nambu and G.Jona-Lasinio, Phys.Rev. {\bf{122}} (1961) 345, 
{\bf{124}} (1961) 246.
  
\item {11.} W.A.Bardeen,C.T.Hill and M.Lindner, Phys. Rev. {\bf{D41}} (1990) 
1647.

\item {12.}  A.Ali Khan, M.Gockeler, T.Horsley, P.E.L. Rakow, G.Schierholz 
and H.Stuben, Phys.Rev. {\bf{D51}}(1995)3751.
\end